\documentclass[reprint, twocolumn, aps, superscriptaddress]{revtex4-1}
\usepackage{amssymb}
\usepackage{amsmath}
\usepackage{epsfig}
\usepackage{color}
\usepackage{graphics, graphicx}
\usepackage{bbold}
\usepackage{psfrag}
\usepackage{mathcomp}
\usepackage{subfigure}
\usepackage{verbatim}
\usepackage{color}
\usepackage[colorlinks, citecolor=blue]{hyperref}

\begin{document}
\title{Majorana edge state in a number-conserving Fermi gas with tunable $p$-wave interaction }
\author{X. Y. Yin}
\affiliation{Department of Physics, The Ohio State University, Columbus, OH 43210, USA}
\author{Tin-Lun Ho}
\affiliation{Department of Physics, The Ohio State University, Columbus, OH 43210, USA}
\affiliation{Institute for Advanced Study, Tsinghua University, Beijing, 100084, China}
\author{Xiaoling Cui}
\email{xlcui@iphy.ac.cn}
\affiliation{Beijing National Laboratory for Condensed Matter Physics, Institute of Physics, Chinese Academy of Sciences, Beijing, 100190, China}

\begin{abstract}
The remarkable properties and potential applications of Majorana fermions have led to considerable efforts in recent years to realize topological matters that host these excitations. For a number-conserving system, there have been a few proposals, using either coupled-chain models or multi-component system with spin-orbit coupling, to create number fluctuation of fermion pairs in achieving Majorana fermion. In this work, we show that Majorana edge
states can occur in a spinless Fermi gas in 1D lattices with tunable $p$-wave interaction. This is facilitated by the conversion between a pair of (open-channel) fermions and a (close-channel) boson, thereby allowing the number fluctuation of fermion pairs in a single chain.
This scheme requires neither spin-orbit coupling nor multi-chain setup and can be implemented easily.
Using the density-matrix-renormalization-group method, we have identified the Majorana phase in a wide range of parameter regime as well as its associated phase transitions. The topological nature of the Majorana phase manifests itself in a strong edge-edge correlation in an open chain that is  robust against disorder, as well as in a non-trivial  winding number in the bulk generated by using  twisted boundary condition. It is also shown that the Majorana phase in this system can be stable against atom losses due to few-body collisions on the same site, and can be easily identified from the fermion momentum distribution. These results pave the way for probing the intriguing Majorana physics in a simple and stable cold atoms system.
\end{abstract}

\maketitle

\section{Introduction}

Majorana fermions, discovered by Majorana in 1937~\cite{Majorana}, has stimulated tremendous research interests over the past decades due to their novel  exchange statistics and promise for topological quantum computation~\cite{RMP1,Alicea}.
A Majorana fermion (or a ``Majorana" for short) is   an equal magnitude superposition of a fermion operator and its adjoint [$\lambda^1 = f^{\dagger}+f$, or $\lambda^2 = i(f^{\dagger}-f) $].   It is a mode of excitation  rather than a particle in the usual sense.  In 2001, Kitaev showed that spinless fermions in a 1D chain coupled to a pairing field will have  Majorana fermions at the ends~\cite{Kitaev}. Efforts to simulate this model in solid state matter have led to the proposal of using
1D semiconducting wires with  spin-orbit coupling (SOC) in contact with a superconductor~\cite{Oreg,Lutchyn}.
Similar proposals have also been made in the cold atom studies by engineering SOC on an attractive Fermi gas~\cite{Jiang, Liu, Wei, Qu, Chen,Wang}.

Since Majorana fermions also emerge in the number conserving systems such as the Pfaffian quantum Hall state~\cite{Moore,Nayak} and Kitaev's honeycomb spin model~\cite{Kitaev2}, there have been questions of whether proximity superconductivity (or lack of number conservation) is necessary for realizing Majaronas in 1D chains.
While a single Majorana excitation can not exist in a number conserving system, the correlation of two Majoranas at different locations  ($i\langle \lambda^{1}_{i}\lambda^{2}_{j}\rangle$) is well defined.  This provides a natural generalization of the presence of Majorana edge modes in a number conserving system, which is defined as a non-zero correlation of the Majoranas at the opposite end of a finite chain, $i\langle \lambda^1_{0}\lambda^2_{N_L}\rangle \neq 0$,  in exactly the same way the Majoranas are correlated in the Kitaev model~\cite{Kitaev}.
In Refs.~\cite{MCheng,Das_Sarma,Fisher},  the authors have studied  coupled 1D chains with interchain pair hopping using bosonization methods and  have concluded the existence of Majorana edge states in these number conserving systems.

Whether Majorana edge states can exist under number conservation is particularly relevant for their realization with cold atoms,  as the latter are number conserving~\cite{Zoller,Diehl,Buchler,Altman,Iemini}. A rigorous proof of their existence was established recently for coupled chain models with inter-chain pair hopping \cite{Zoller,Diehl,Buchler}, and
for a single chain four-component fermion system with SOC and spin-exchange interaction ~\cite{Iemini} which has the similar pair hopping physics as in other coupled chain models.  All these  studies suggest that the number fluctuation of fermion pairs in a single chain is the key to the emergence of Majoranas in number conserving systems.

\begin{figure}[t]
\includegraphics[angle=0,width=65mm]{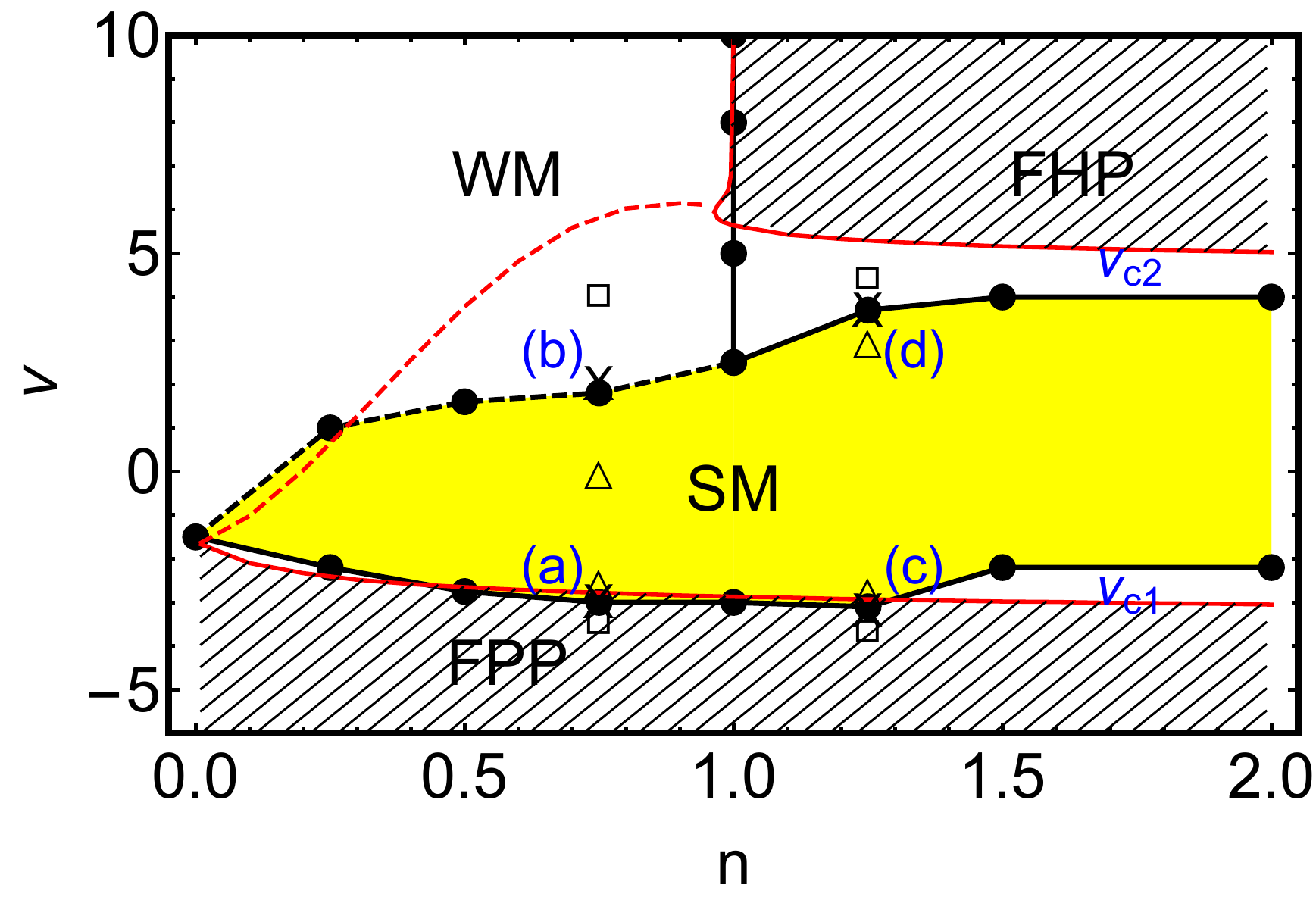}
\caption{Phase diagram in terms of the total filling $n=N/N_L$ and the boson detuning $\nu$ at a fixed inter-channel coupling $g=1$. The energy unit is $t_f$. The red lines are mean field phase boundaries.  The dots connected by black lines are phase boundaries based on DMRG calculations. The dashed lines mark the crossover from the strong Majorana ($SM$) to the weak Majorana ($WM$) region,  rather than the sharp phase transition.
The yellow color highlights the $SM$ region with strong edge-edge correlation from DMRG calculations. The shaded area shows non-topological superfluid regime from mean-field theory; the upper and lower areas are the superfluids made up of a gas of fermion hole pairs ($FHP$) and fermion particle pairs ($FPP$), respectively, in addition to a Bose condensate.  The sets of squares, crosses, and triangles labelled (a), (b), (c), and (d) correspond to curves plotted in Fig.~\ref{fig_G}(a) to (d).}
\label{fig1}
\end{figure}

In this paper, we propose a much simpler scheme for realizing Majorana edge states that makes use of alkali fermions in a single chain without SOC -- by simply tuning a single component Fermi gas in a 1D chain to its $p$-wave resonance. 
The systematic derivation of this model is given in Ref.~\cite{Cui}.
In the two-channel description of $p$-wave resonance,  two fermions at  neighboring lattice sites can convert to a ``close channel" boson in one of the two sites, thereby causing the number fluctuation of fermion pairs in the  chain. The close channel bosons play  the role of proximity superconductor in the Kitaev model, except that they are now quantum mechanical objects. In this model, neither the number of  fermion  $N_{f}$ nor the number of boson $N_b$ is conserved. However, the total number $N=2N_{b}+N_{f}$ is [See Eq.~(\ref{eq_ham_full})]. An effective fermion-molecule conversion model was also proposed previously using the laser-assisted pair tunneling\cite{Sylvain}. Here through exact numerical calculations, we confirm the Majorana ground state with strong edge-edge correlations in a broad range of paramters : filling factor, boson detuning,  and inter-channel coupling (See yellow regions in Fig.~\ref{fig1} and Fig.~\ref{fig_g}).
It is useful to contrast our model with the single-channel model that only consists of spinless fermions with neighboring-site attractions. While both models are number conserving and their mean field theories share the same structure,  our exact calculations show that only the  resonant two-channel model exhibits strong Majorana correlations. This shows again the essential role played by the number fluctuation of fermion pairs in the chain, which is absent in the single-channel attractive Fermi gas.

Experimentally,  a major obstacle for exploring $p$-wave effect in cold atoms is the severe atom loss, as observed in a 3D Fermi gas~\cite{JILA_K40, Li6_1, Li6_2, Toronto_K40}. Recent studies have suggested that the $p$-wave system could be more stable against three-body loss if confined in the quasi-1D geometry~\cite{Cui1,Cui2}. In our system, an additional optical lattice is applied along the 1D tube and the space is further discretized. In this case, the atom loss comes from the possibility of finding pairs of boson-fermion or boson-boson at the same site  and their collisions at close proximity.
Here we show that within the two-channel $p$-wave model, the probabilities of finding such pairs are very low in a large region of parameter space for the Majorana phase. This provides  promising prospects to realize the Majorana phase and to perform studies with a wide range of cold atom techniques.

\section{Model}

Our model is~\cite{footnote_model}
\begin{eqnarray}
H=&\sum_{j}\left( -t_b b_j^\dagger b_{j+1}+\text{h.c.} \right)
+\sum_{j}\left(-t_f f_j^\dagger f_{j+1}+\text{h.c.} \right) \nonumber\\
&+\nu\sum_{j}  b_j^\dagger b_{j}+g \sum_{j} \left[ b_j^\dagger(f_{j-1}f_j+f_{j}f_{j+1})+\text{h.c.} \right], \label{eq_ham_full}
\end{eqnarray}
where $b^\dagger_{j}$ and $f^\dagger_{j}$ create a closed-channel boson and an open-channel fermion at site $j$,  with nearest-neighbor hopping $t_b$ and $t_f$ respectively; $g$ is the ($p$-wave) inter-channel coupling and $\nu$ is the boson detuning.
As $\nu$ approaches zero, conversion between bosons and fermion pairs for given $g$ will become more frequent due to their energy match.
The number of boson ($N_b=\sum_i b^{\dag}_ib_i$) and  fermion  ($N_f=\sum_i  f^{\dag}_if_i$) are not  separately conserved, but the  sum $N=2N_b+N_f$ is.  Since bosons are heavier than fermions, $t_b$ is smaller than $t_f$. In this paper, we will take $t_b=0.2t_f$ and use $t_f$ as the energy unit.

If $b_j$ in Eq.~(\ref{eq_ham_full}) is replaced by a $c$-number,  as in  mean field approach,
Eq.~(\ref{eq_ham_full}) reduces to the Kitaev model~\cite{Kitaev}, which has a Majorana phase. The question is whether this phase will survive the quantum fluctuation of the bosons,  such that the ground state of Eq.~(\ref{eq_ham_full}) will have the same edge-edge Majorana correlations as in the Kitaev model.  This is to be answered in this work.

\section{Mean-field analysis}

We first carry out mean-field analysis to gain a qualitative understanding of  the  problem.
Assuming the bosons fully condense at zero momentum, i.e., $\langle b_{k=0}\rangle=\sqrt{N_b}$,
the Hamiltonian $\Omega=H-\mu N$ in  grand canonical ensemble can be written as \begin{eqnarray}
\label{eq_ham_mf}
&\Omega=\sum_{k}\left[-2t_f \cos(k d)-\mu\right]f_k^\dagger f_k +N_b(-2t_b +\nu-2\mu) \nonumber\\
&+(\Delta/2)\sum_k \left[ -i
 \sin(kd) f_k f_{-k}+\text{h.c.} \right], \label{h_mf}
\end{eqnarray}
here $d$ is the lattice spacing; $\Delta=4g\sqrt{n_b}$ with $n_b=N_b/N_L$ the boson filling and $N_L$ the number of lattice sites.  Equation~(\ref{h_mf}) can be written as
\begin{equation}
\frac{\Omega}{N_L} = \frac{1}{N_L} \sum_{k>0} \left( F^{\dagger}_{\mu}(k)  {\cal H}_{\mu \nu}(k) F_{\nu} (k)+\xi_k\right)+\frac{|\Delta|^2}{16g^2}(-2t_b+\nu-2\mu),
\end{equation}
with $\xi_{k}  = - 2t_f \cos(kd)-\mu$, $F=(f_{k},f_{-k}^{\dagger})^T$ and
\begin{equation}
{\cal H}(k) =  \left(  \begin{array}{cc}  \xi_{k}    &  -i \Delta \sin(kd)
 \\ i  \Delta \sin(kd) & -  \xi_{k}  \end{array} \right).
\label{calH}
\end{equation}

Minimizing the ground state energy of Eq.~(\ref{h_mf}) with respect to $\Delta$ and imposing the number constraint $N=2N_{b} + \sum_{k}\langle f^{\dagger}_{k}f^{}_{k}\rangle$, we  obtain the gap equation and number equations as:
\begin{eqnarray}
\frac{-2t_b+\nu-2\mu}{16g^2} &=& \frac{1}{N_L}\sum_{k>0}\frac{[\sin(kd)]^2}{2E_k}\label{gap_eq},
\end{eqnarray}
and
\begin{eqnarray}
n(1-c_b)&=&\frac{1}{N_L} \sum_{k>0} \left( 1-\frac{\xi_k}{E_k} \right),
\label{number_eq}
\end{eqnarray}
where $E_k=\sqrt{\xi^{2}_{k} + (\Delta\sin(kd))^2 }$ is the
excitation spectrum, $c_b=2n_b/n$ is the boson fraction,  and
$n=N/N_L$ is the total filling. Unlike the Kitaev model where the
$p$-wave pairing and the chemical potential are both external
inputs~\cite{Kitaev}, here $\Delta$ and $\mu$ are determined
self-consistently for given $g$, $\nu$ and $n$.

Since the mean-field Hamiltonian Eq.~(\ref{h_mf}) is in the form of
the Kitaev model, the Majorana phase lies in the region $|\mu|<2t_f$
as pointed out in Ref.~\cite{Kitaev}. The topological character of the ground state is specified by the Berry Phase of the ground state $\chi(k)$ of ${\cal H}(k)$, integrated over the Brillouin Zone, i.e. 
$\Delta \Phi \equiv  i  \int^{\pi/d}_{-\pi/d}  {\rm d}k   \chi^{\dagger}(k) \partial_{k} \chi(k) $.  Writing ${\cal H}(k)$ as 
${\cal H}(k) = {\bf h}(k)\cdot \vec{\sigma}$, where
\begin{equation}  
{\bf h}(k) = \xi_{k} \hat{\bf z}    + \Delta\sin(kd) \hat{\bf y} \equiv E_{k} \left( {\rm cos}\theta_{k} \hat{\bf z} +  {\rm sin}\theta_{k}\hat{\bf y}\right),  \label{h_k}
\end{equation}
we have 
\begin{equation}
\chi(k) = \left( \begin{array}{c}  i  {\rm sin}(\theta_{k}/2) \\   {\rm cos}(\theta_{k}/2) \end{array}\right) e^{i\gamma(k)}.
\end{equation}
Here the phase factor $e^{i\gamma(k)}$ is to keep  $\chi(k)$ periodic,
\begin{equation}
\chi(k) = \chi(k + 2\pi/d).
\label{BC} \end{equation}
As seen from Eq.~(\ref{h_k}), the tip of  ${\bf h}(k)$ traces out a closed curve in the $yz$-plane as $k$ varies from $-\pi/d$ to $\pi/d$. For $|\mu |< 2t_{f}$, this curve encloses the origin, and we have $\theta(-\pi/d) = 0$, and $\theta(\pi/d)= 2\pi$. In order to ensure the periodicity of $\chi(k)$ (Eq.~(\ref{BC})), $\gamma$ has to satisfy the condition $\gamma(-\pi/d) - \gamma(\pi/d)=\pi$, which gives the Berry phase $\Delta \Phi=\pi$. This is the topologically non-trivial (Majorana) phase. Otherwise for $|\mu |> 2t_{f}$, the curve of ${\bf h}(k)$ does not enclose the origin, and we have $\theta(\pi/d) = \theta(-\pi/d),\ \gamma(\pi/d) = \gamma(-\pi/d)$, and thus $\Delta \Phi=0$. This is the topologically trivial phase.

After $\Delta$ and $\mu$ are determined
self-consistently, we obtain the mean field phase diagram in Fig.~\ref{fig1} over a broad
range of $\nu$ and $n$ for a fixed coupling $g=1$. There are three
superfluid phases (all with $\Delta\neq 0$): a topological Majorana
phase  $M=SM + WM$, consisting a ``strong Majorana" region $SM$ and
a ``weak Majorana" region $WM$, and two non-topological superfluid phases
$FHP$ (fermion hole pairs) and  $FPP$ (fermion particle
pairs).  The reasons of the nomenclature will be explained further later.

The phase boundaries separating the Majorana phase $M=SM+WM$ and the $FHP$  ($FPP$) phase is determined by the condition $\mu=2t_f$ ($\mu=-2t_f$), as shown by red lines in Fig.~\ref{fig1}.  Our numerical solutions of Eqs.~(\ref{gap_eq}) and (\ref{number_eq}) show that the upper boundary ($\mu=2t_f$) takes a sudden upward turn at $n=1$. The red dashed line in Fig.~\ref{fig1} marks a crossover from SM to WM regime, when the boson fraction $c_b$ continuously decays to a small value $0.01$. Later, we show from exact numerical calculations that these two regions can be distinguished from the behavior of edge-edge Majorana correlations. 

The $FPP$ phase exists in sufficiently negative $\nu$, where the system is mostly in the Bose condensate, with a dilute gas of fermion particles forming pair superfluid.  In contrast, the $FHP$ phases exists for $n>1$ and for sufficiently high $\nu$. In this regime, the fermion occupation is more favored than bosons, leading to the nearly full fermion filling $n_f\sim1$ with a small fraction of fermion holes. Since the fermion superfluidity essentially relies on the number fluctuations, it can be viewed as the pairing of fermion holes.


\section{DMRG analysis}
We have calculated the ground state properties of the Hamiltonian  [Eq.~(\ref{eq_ham_full})] using  density-matrix-renormalization-group (DMRG) method~\cite{alps1, alps2}. The calculations are done with maximum $800$ truncated states and $30$ sweeps, and the truncation error is $10^{-8}$.   Since we consider the filling regime with $n\le 2$, we have set the truncated number of bosons at each site as up to four in our simulation.

\subsection{Identification of Majorana phase and Majorana edge state}

In this section, we use  DMRG to work out the phase diagram of Eq.~(\ref{eq_ham_full})  [(I) and (II) below], and then examine in (III) the presence of Bose condensation (i.e superlfuid nature) in different phases, followed by a study in (IV) of the Majorana correlation in these phases which determines the presence of Majorana edge modes. The section will be ended by a discussion of the momentum distribution which shows the distinct signature of the Majorana phase.

{\it (I) Entanglement entropy.}  It is known that a thermodynamic phase transition is reflected in
a singularity of the entanglement entropy at the transition point~\cite{amico}.
Given the many-body ground state $|\psi\rangle$, the reduced density matrix can be written as $\rho_L=\text{Tr}_R |\psi\rangle\langle\psi|$, with $L$, $R$ denoting the left and right half of the lattice. Its eigenvalues $\{\lambda_{\alpha}\}$ determine the entanglement entropy $S=-\sum_{\alpha}\lambda_{\alpha}\ln \lambda_{\alpha}$.

\begin{figure}[htbp]
\includegraphics[angle=0,width=65mm]{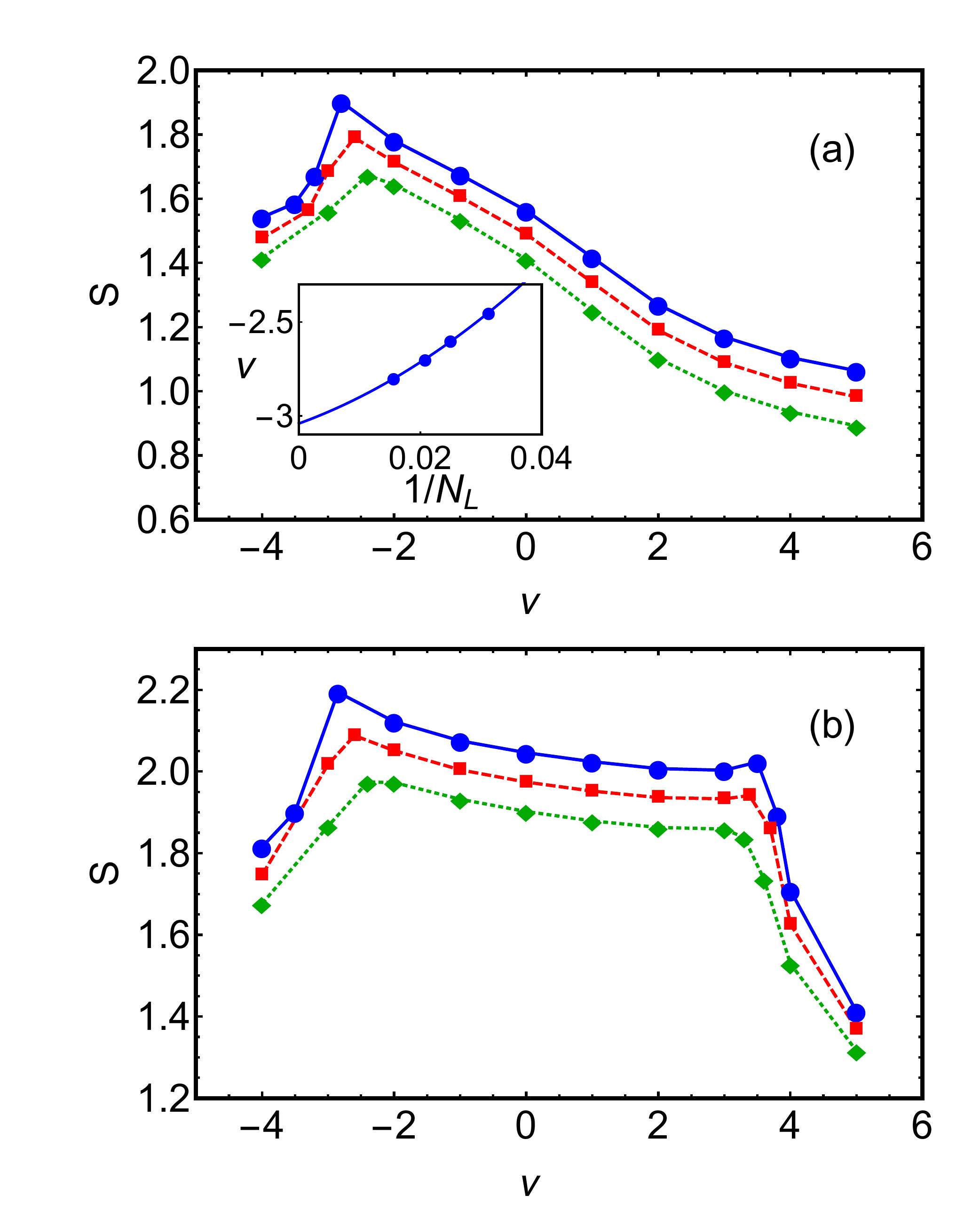}
\caption{Entanglement entropy $S$ as functions of $\nu$ for (a) $n=0.75$ and (b) $n=1.25$ with different system sizes $N_{L}$. Green diamonds, red squares and blue circles correspond to $N_L=24, 40$ and $64$, respectively. Here $g=1$. Inset in (a) shows the extrapolation to infinite system for the case of $n=0.75$.}
\label{fig_S}
\end{figure}
We find that the behavior of $S$ as a function of detuning $\nu$ depends on the filling $n$.
For $n<1$, we find one cusp in $S$ at $\nu=\nu_{c1}$ (Fig.~\ref{fig_S}(a)); while for $n>1$, we find two cusps in $S$ at $\nu_{c1},\nu_{c2}$ ($\nu_{c1}<\nu_{c2}$, see Fig.~\ref{fig_S}(b)).    This indicates one or two transitions by tuning $\nu$. Repeating the calculation for different sample sizes,  we have  obtained the estimate of the critical values of $\nu_{c1},\nu_{c2}$  for infinite systems through extrapolation (see the inset of Fig.~\ref{fig_S}(a)). These phase boundaries are shown by dots connected by solid lines in the phase diagram in Fig.~\ref{fig1}.  Here we have chosen $g=1$, and we find these phase boundaries are  close to those obtained by mean field theory. Accordingly, we adopt the same nomenclature as in mean field theory for the phases obtained from DMRG.  \\

\begin{figure}[htbp]
\includegraphics[angle=0,width=85mm]{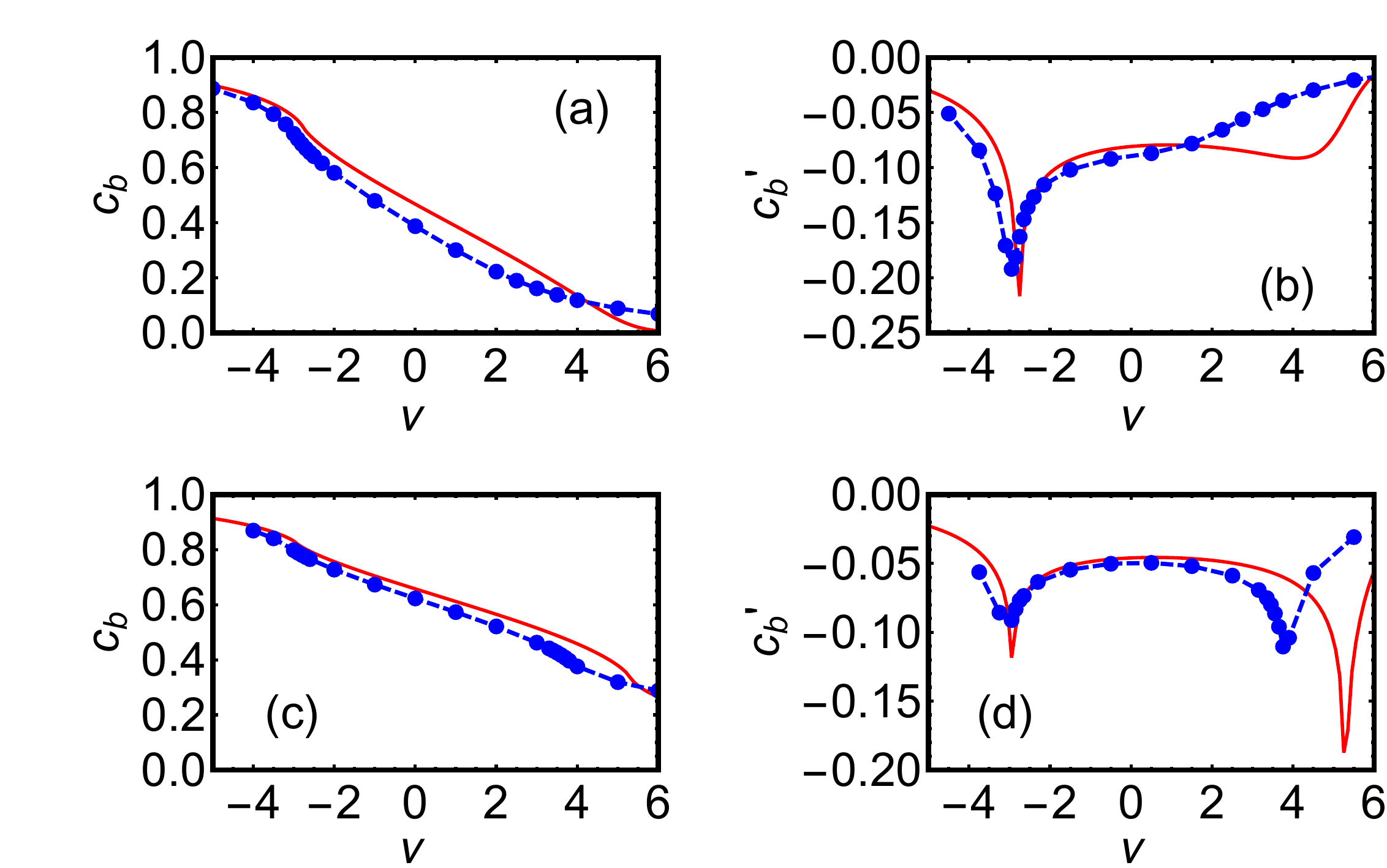}
\caption{Boson fraction $c_b$ (a, c) and its variation $c_b'$ (b, d) as functions of $\nu$.
Panel (a, b) and (c, d) are for  fillings $n=0.75$  and  $n=1.25$  with $N_L=64$ and $g=1$.
Dots and dashed blue lines show the results determined by DMRG. Solid red lines show the results of mean-field predictions.}
\label{fig_n}
\end{figure}

{\it (II) Boson fraction and its variation.}  We have also verified the phase boundaries by calculating the boson fraction $c_b$ and its variation $c_b' \equiv \partial c_b/\partial \nu$  as a function of $\nu$.
The results are shown in Fig.~\ref{fig_n}.  One sees that  while $c_b$ is continuous in $\nu$,  $c_b'$ has one or two sharp cusps depending on whether the filling $n<1$ or $n>1$. The locations of these cusps are consistent with those obtained in (I), thus confirming the phase transitions discussed in (I).
 In addition, the singularity in $c_b'$ shows that the transitions are of the second order. In Fig.~\ref{fig_n}, we also compare with the mean-field predictions (thin red lines) and find qualitative agreement.\\

{\it (III)  Condensation of bosons and fermion pairs: } The criterion of Bose condensation was generalized to interacting system by Penrose and Onsager ~\cite{PO}. It also applies to finite number systems.  The Penrose-Onsager criterion makes use of the property of the single particle density matrix  $\rho_{ij}  \equiv \langle b^{\dagger}_{i} b_{j}\rangle$ evaluated for the state of interest.  The state is Bose condensed if $\rho$ has a {\em single} maximum eigenvalue $\zeta_{1}$  of order $N_b$, while the ratios  $\zeta_{\beta}/\zeta_{1}$  for all other eigenvalues $\zeta_{\beta}$ are much less than 1, and tend to zero as $N_b$ goes to infinity. The eigenfunction associated with $\zeta_{1}$ is referred to as the condensate wavefunction. Similarly, one can also define a fermion pair correlation matrix $\eta_{ij}\equiv \langle f^{\dagger}_{i+1}f^{\dagger}_{i}  f^{}_{j} f^{}_{j+1}\rangle$. Condensation of fermion pairs [as characterized by C.N. Yang as off-diagonal long-range order, see Ref.~\cite{Yang}] corresponds to a single large eigenvalue of $\eta$ of order $N$, as in the bosonic case.  

\begin{figure}[htbp]
\includegraphics[angle=0,width=85mm]{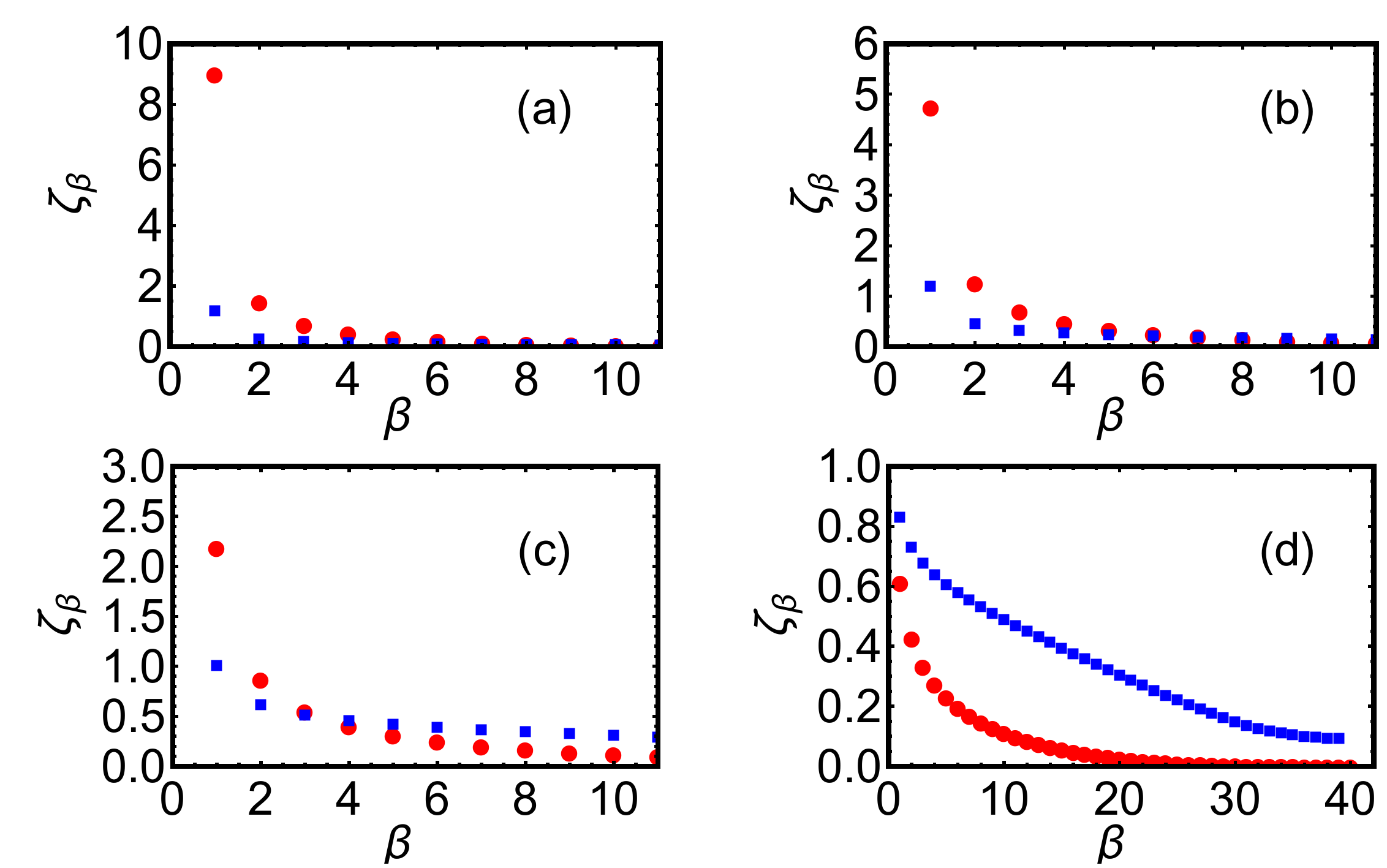}
\caption{Red dots show eigenvalues $\zeta_{\beta}$ of single particle density matrix  $\rho_{ij}  \equiv \langle b^{\dagger}_{i} b_{j}\rangle$. Blue squares show eigenvalues of  fermion pair correlation $\eta_{ij}\equiv \langle f^{\dagger}_{i+1}f^{\dagger}_{i}  f^{}_{j} f^{}_{j+1}\rangle$. The eigenvalues are sorted from the largest to the smallest. 
(a), (b), (c), and (d) show the cases with $\nu=-4$, $-2$, $0$, and $2$, respectively, for $n=0.75$. They correspond to FPP, SM, SM, and WM states in Fig.~\ref{fig1}, respectively.
Total number of bosons $N_b$ in (a), (b), (c), and (d) are 12.57, 8.80, 5.82, and 3.31, respectively. 
Here $g=1$ and $N_L=40$.}
\label{PO}
\end{figure}

Using DMRG, we have calculated the eigenvalues of the matrices $\rho_{ij}$ and $\eta_{ij}$ for all the ground states in different phases in  Fig.~\ref{fig1}.  To illustrate the Bose condensation, we have shown our results in Fig.~\ref{PO}(a)-(d) as one increases the detuning from $-4$ to $2$ at $n=0.75$.  This path takes one from the $FPP$ phase to the $SM$ region and then to the $WM$ region. We see from Fig.~\ref{PO}(a)-(c) that both  $FPP$ and  $SM$ phases have a distinct large eigenvalue, whereas the large eigenvalue  gradually disappears into a continuous  (power-law like) distribution of eigenvalues  in the $WM$ region, see Fig.~\ref{PO}(d).   In contrast, the fermion pair correlation $\eta$  does not show a  large distinct eigenvalue in all cases, and appears to be power-law like.  The situation at the $FHP$ phase is similar to that of the $FPP$ phase, i.e. there is a distinct eigenvalue for the boson density matrix, and a power-law like distribution for the fermion pair distribution.  \\

{\it (IV) Edge-edge Majorana correlation.} The emergence of Majorana fermion is associated with long range edge-edge correlations. In the Kitaev chain model~\cite{Kitaev}, one defines two sets of Majorana fermion operators $\lambda^1_{i}=f^{\dag}_i+f_i$ and $\lambda^2_{i}=i(f^{\dag}_i-f_i)$, and the Majorana phase can be characterized by the order parameter ${\cal O}\equiv i\langle \lambda^1_1 \lambda^2_{N_L} \rangle$, which directly manifests the correlation between two Majorana modes at different edges~\cite{Kitaev, Zoller,Diehl,Buchler,Iemini}.
For a number conserving system, $|{\cal O}|$ is directly reduced to the edge-edge correlation function $G(1,N_L)$, where $G$ is defined as:
\begin{equation}
G(i,j)\equiv  | \langle f^\dagger_i f_{j} + h.c.\rangle|.
\end{equation}

\begin{figure}[h]
\includegraphics[angle=0,width=85mm]{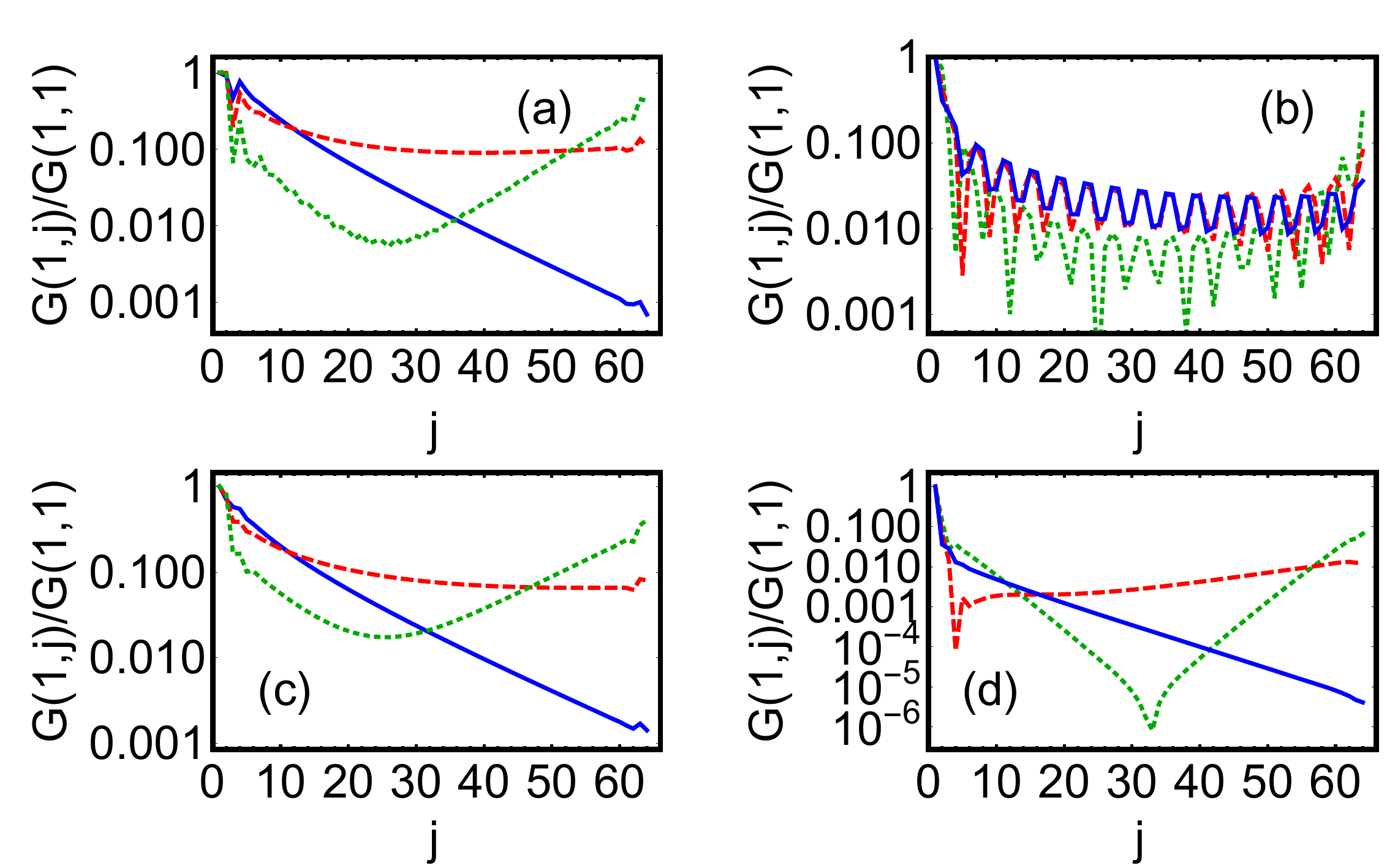}
\caption{The behavior of the correlation function $G(1,j)$ in log scale as one passes through the boundaries as shown in Fig.~\ref{fig1}.   (a,b) are for filling $n=0.75$, and (c,d) are for $n=1.25$. The blue solid, red dashed and green dotted lines from (a) to (d) correspond to ($\nu=-3.5, -3, -2.5$), ( $\nu=4, 2, 0$), ($\nu=-3.7, -3.2, -2.7$), and ($\nu=4.4, 3.7, 3$).
In (a,c,d), when cross the phase boundaries into SM phase, $G(1,j)$ shows strong revival as $j$ approaches the other edge $j\rightarrow N_L$. In (b), when passing from the $WM$ into the $SM$ region, $G(1,j)$ changes slowly even within a large range of $\nu$. In $WM$ region, $G(1,N_L)$ dropped to a small but non-zero value ($G(1,N_L)/G(1,1)<10\%$). Here we take $g=1$ and $N_L=64$.}
\label{fig_G}
\end{figure}

 In Fig.~\ref{fig_G}, we show the behavior of correlation function $G(1,j)$ as the system passes through different boundaries  marked  (a) to (d) in Fig.~\ref{fig1}. For filling $n>1$, as the system enters the  $SM$ phase through the phase boundary $\nu_{c1}$ from the $FPP$ phase below, or through the phase boundary $\nu_{c2}$ from the $FHP$ phase above,
$G(1,j)$ immediately shows strong revival as $j$ approaches the other edge, which is the hallmark of Majorana edge states. See
Fig.~\ref{fig_G}(c) and Fig.~\ref{fig_G}(d).
In contrast, in the $FPP$ phase ($\nu<\nu_{c1}$) or $FHP$ phase ( $\nu>\nu_{c2}$), $G(1,j)$ decreases exponentially fast as $j$ increases, indicating the absence of Majorana edge correlations.

For $n<1$, similar revival behavior also shows up as $\nu$ across the lower boundary $\nu_{c1}$ (see Fig.~\ref{fig_G}(a)) . Here, within a small range of $\nu$ from $-3.5$ to $-2.5$, the edge-edge correlation $G(1,N_L)$ (scaled by $G(1,1)$) increases from $10^{-3}$ to as large as $0.6$.
Continuously increasing $\nu$, $G(1,j)$ crossovers to a different behavior.  It decreases slowly without strong revival but reaching a small yet non-zero value as $j$ approaches the other edge, see Fig.~\ref{fig_G}(b). Specifically, for a large range of $\nu$ from $0$ to $4$, $G(1,N_L)$ decreases from $0.2$ to $0.05$.
Further increasing $\nu$, $G(1,N_L)$ becomes even smaller but still finite (not exponentially small as in the non-Majorana phases $FPP$ and $FHP$). 
In this sense, in Fig.~\ref{fig1} we draw a dashed line for filling $n < 1$ to show the crossover from the strong Majorana (SM) to weak Majorana (WM) when $G(1,N_L)$ decays to 10$\%$ of the onsite $G(1,1)$. In this way we highlight the SM region with yellow color in Fig.~\ref{fig1}, where one can find strong Majorana edge-edge correlations.\\

\begin{figure}[htbp]
\includegraphics[angle=0,width=65mm]{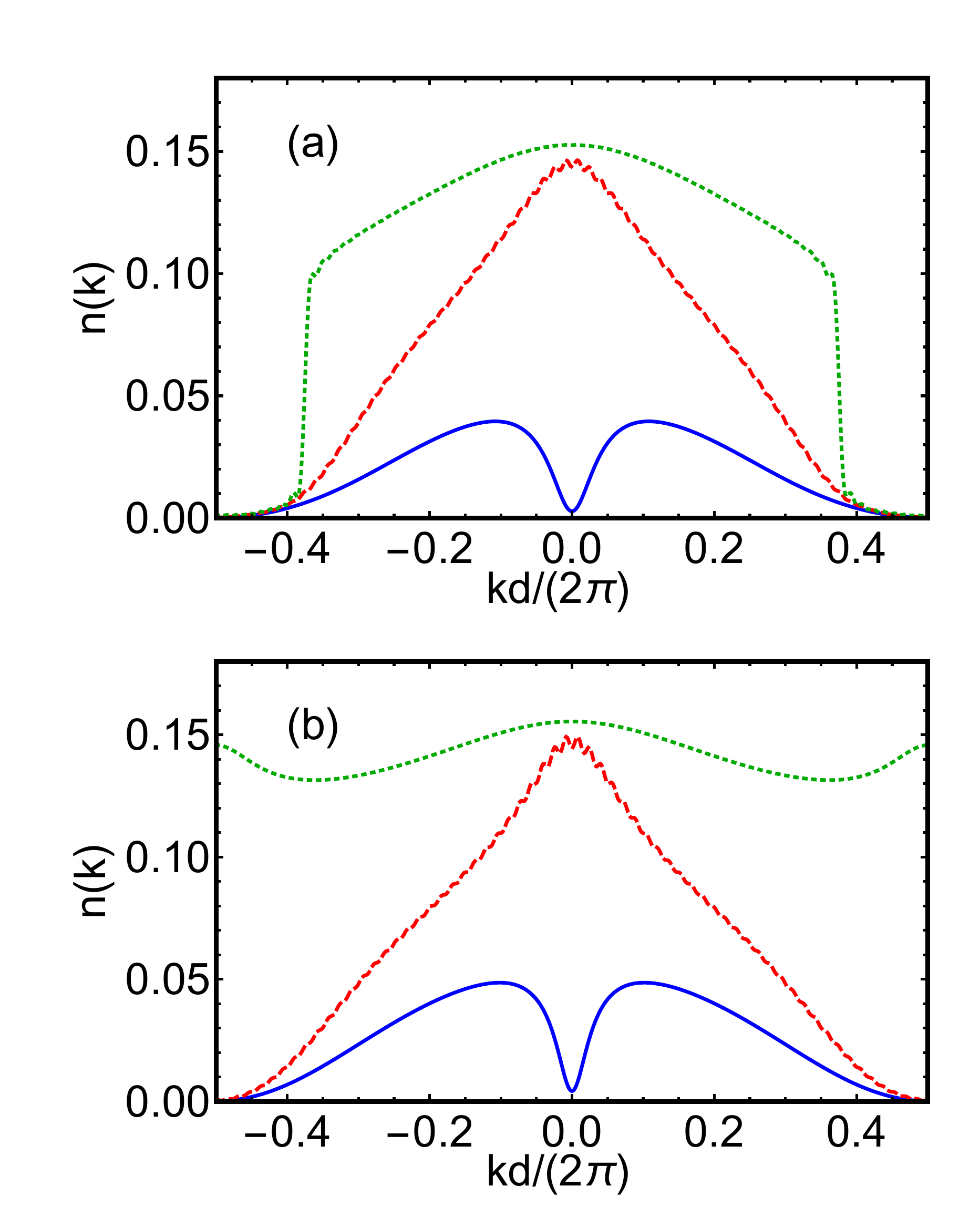}
\caption{Fermion distribution in momentum space for different $\nu$ at filling
(a) $n=0.75$ and (b) $n=1.25$.
Blue solid, red dashed, and green dotted lines in (a) show $n(k)$ for $\nu=-4$, $-1$, and $3$, corresponding to FPP, SM, and WM phases, respectively.
Blue solid, red dashed, and green dotted lines in (b) show $n(k)$ for $\nu=-4$, $-1$, and $6$, corresponding to FPP, SM, and FHP phases, respectively.
Here $g=1$ and $N_L=64$. }
\label{fig7}
\end{figure}

{\it (V)  Momentum distribution: } The behavior of correlation function in spatial space ($ \sim\langle f^{\dagger}_{i} f_{j}\rangle$) directly determines its Fourier transformation, i.e., the momentum distribution of fermions $n(k) = \langle f^{\dagger}_{k} f_{k}\rangle$ that can be easily measured in cold atoms experiments. 
In Fig.~\ref{fig7}, we show $n(k)$ from DMRG calculation for three typical values of $\nu$ at filling $n=0.75$ and $1.25$. We can see that in both cases, when $\nu$ stays in the SM region(red dashed line in Fig.~\ref{fig7}), $n(k)$ features a distinct peak at $k=0$ while decays to zero at the Brillouin edge $k=\pi/d$. We have checked that such property of $n(k)$ holds true for all SM states in Fig.~\ref{fig1}.
In contrast, $n(k)$ is roughly a constant in the $FHP$ phase, and has a hole at $k=0$ in the $FPP$ phase. $n(k)$ for WM state share similar structure as the SM case, while its peak at $k=0$ is not as sharp as the latter.   

All these  behaviors can be understood from the mean-field theory, where we have the analytical expression $n(k) = [1 + (2t_f \cos(kd) +\mu)/E_k]/(2N_L)$ (see Eq.~(\ref{number_eq})). It is easily seen that within the Majorana region ($|\mu|<2t_f$), $n(k)$ is the largest at $k=0$ while it is zero at $k=\pm \pi/d$. In the  $FHP$ phase, the detuning $\nu$ is large and positive, which leads to a large and positive $\mu(>2t_f)$ in the mean field theory and a finite $n(k)$ at both $k=0$ and $k=\pm \pi/d$. Such detuning makes it costly to create bosons, making the system more free fermion like.
In the $FPP$ phase, $\nu$ is large and negative, forcing most of the particles into  the $k=0$ Bose condensate, and leads to a large and negative $\mu(<-2t_f)$ in mean field theory.  This leads to $n(k)=0$ at both $k=0$ and $k=\pm \pi/d$. In particular, at small $k$, we have $n(k) = |\Delta\sin(kd)/(2\mu)|^2 \sim k^2 $.

Given the distinct $n(k)$ for different phases, they can be used as experimental indicators of various phases in current system.  Note, however, that $n(k)$ cannot be directly related to the edge mode in Majorana physics, because it contains a large contribution from the bulk. This is evidenced by the fact that the behaviors of $n(k)$ are well accounted for by the results of mean field theory. \\

\subsection{Robust topological features of the strong Majorana phase}

In this section, we will show that the strong Majorana (SM) phase has robust topological features,  in that  the edge-edge correlation survives from disorders, and it can host a non-trivial winding number in the bulk system. We will also show that the condensation of bosons and the
edge-edge correlation remain strong for increasing system size.

\begin{figure}[htbp]
\includegraphics[angle=0,width=65mm]{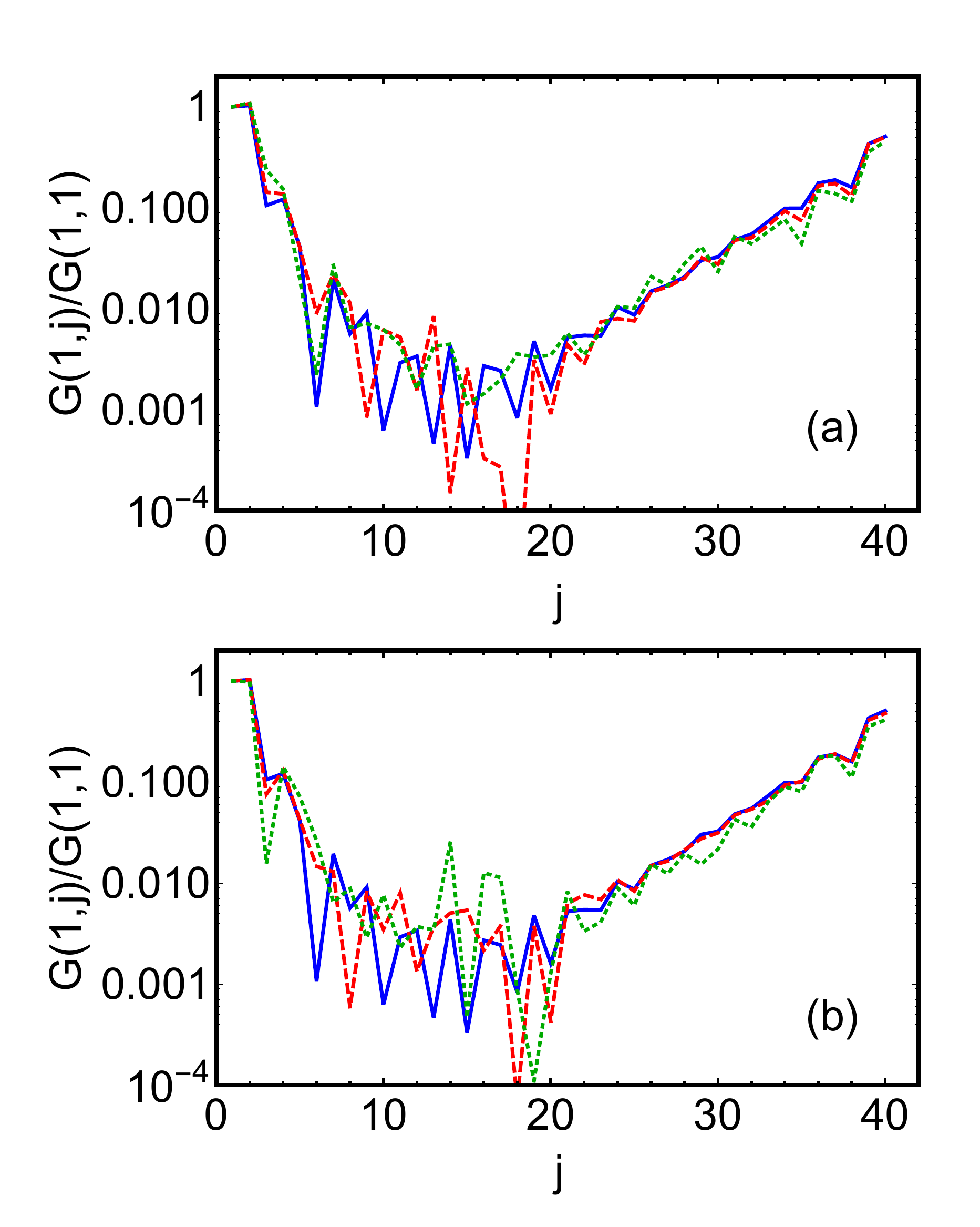}
\caption{The correlation function $G(1,j)$ in the presence of disorder in hopping (a) and disorder in inter-channel coupling (b). Different disorder strengths $D=0$(blue solid), $0.1$(red dashed) and $0.25$(green dotted) are shown. Here $n=0.75$, $g=1$,  $\nu=-2$, and the system stays in strong Majorana (SM) phase. }
\label{fig_disorder}
\end{figure}

Firstly, we study the robustness of edge-edge correlation in the presence of disorder. Here we impose disorder either in the fermion hopping term ($t_f$) or in the coupling term ($g$) in the Hamiltonian (\ref{eq_ham_full}), and carry out DMRG simulations with OBC. Specifically, $t$ or $g$ are now site-dependent, $t_f\rightarrow t_f+D\delta_j$ or $g\rightarrow g+D\delta_j$ ($j$ is site index); here $\delta_j\in(-1,1]$ is a random number, and $D$ is the strength of disorder. In Fig.~\ref{fig_disorder}, we show the behavior of correlation function $G(1,j)$ for a typical SM ground state with different disorders in $t_f$ [Fig.~\ref{fig_disorder}(a)] and $g$ [Fig.~\ref{fig_disorder}(b)]. We see that small disorder cannot change the strong revival character of $G(1,j)$  even for $D$ reaching $0.25$. This shows the edge modes are robust against a fairly large amount of external perturbations.

\begin{figure}[htbp]
\includegraphics[angle=0,width=65mm]{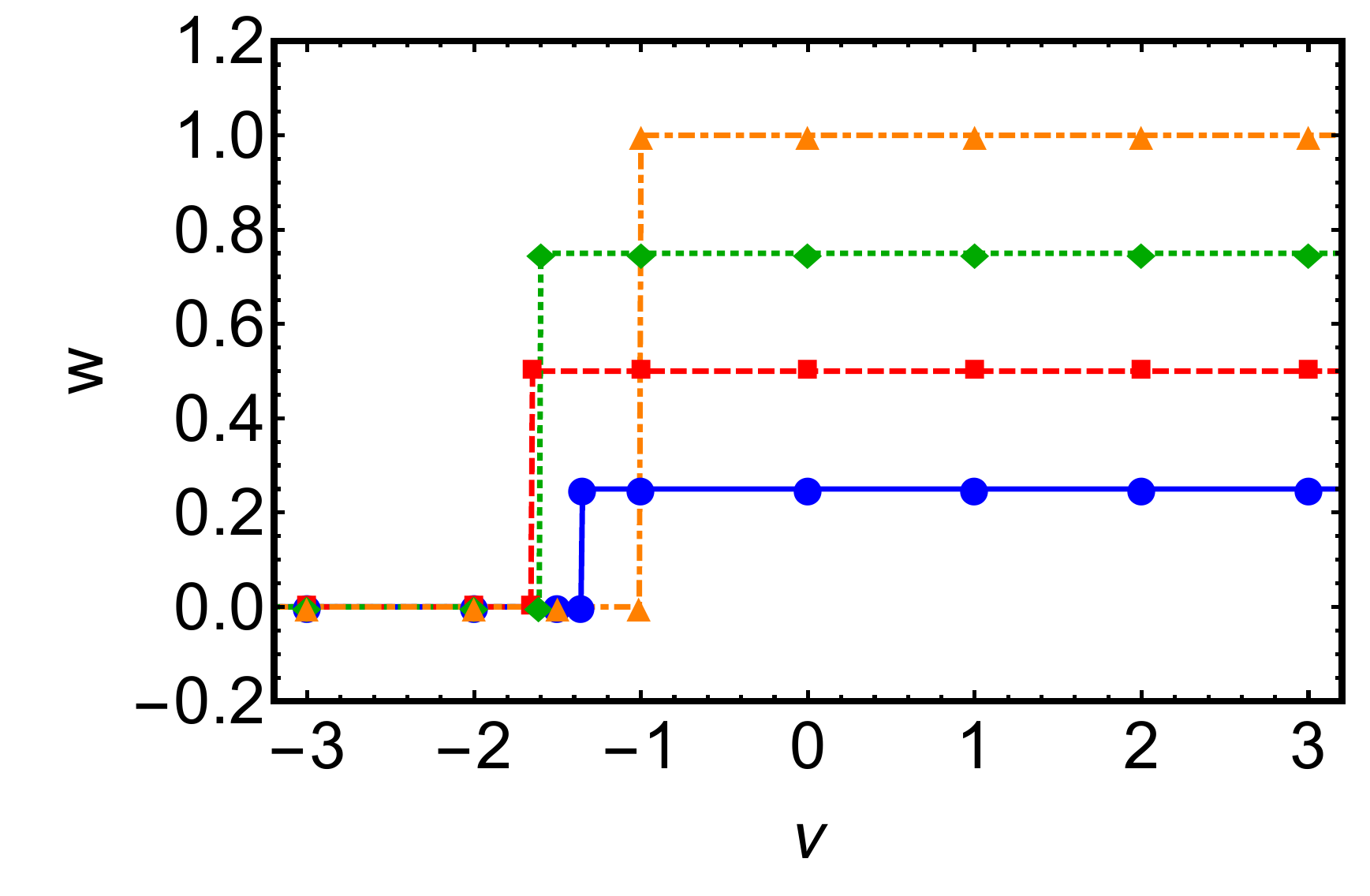}
\caption{Winding number [Eq.~(\ref{w})] as functions of boson detuning for small size systems with twisted boundary. Blue circles, red squares, green diamonds and orange triangles are respectively with $(n,N_L)=(0.25,8),\ (0.5,8), \ (0.75, 8)$, and $(1,6)$. }
\label{fig_w}
\end{figure}

Secondly, in order to demonstrate the topological feature of the bulk system, we impose a twisted phase  boundary condition and calculated the resulting winding number. Specifically, we turn on the fermion hopping between the two edges ($i=1$ and $L$) as $t_{1L}^*=t_{L1}=t_f e^{i\theta}$, where $\theta\in(0,2\pi]$ is the twist phase. In the single-particle picture, this corresponds to shifting the momentum as $k\rightarrow k+\theta/N$, and when $\theta$ varies from $0$ to $2\pi$, the momentum basis $\{ k\}$ returns to itself and completes a closed loop, so as the Hamiltonian $H$. For the interacting many-body state, we calculate the winding number of the form
\begin{equation}
w=i\int_0^{2\pi} d\theta \langle \Psi_{\theta}| \partial_{\theta}| \Psi_{\theta}\rangle/\pi, \label{w}
\end{equation}
with $\Psi_{\theta}$ is the ground state with twisted phase $\theta$. In Fig.~\ref{fig_w} we show  $w$ as a function of detuning $\nu$ for several fillings by exactly diagonalizing  small size systems. We find that given the filling factor $n$, by increasing the detuning $\nu$ to drive the system from FPP to SM phase, $w$ will have a sudden jump from $0$ to a finite value ($=n\pi$) at a critical $\nu_c$ (in thermodynamic limit $\nu_c$ is expected to recover the lower boundary as shown in Fig.~\ref{fig1}). This signifies a topological transition between the two phases. The finite $w$ continues to the WM phase when further increasing $\nu$.  

Note that here $w$ depends on the filling factor, instead of a constant ($\pi$) Berry phase in the mean-field analysis (see section III). 
This difference can be attributed to different ways in introducing a closed path in parameter space. Specifically, in the mean-field analysis, the closed path is completed by moving $k$ through the entire Brillouin Zone, which in the single-particle picture corresponds to fermions occupying a Fermi-sea at full filling $n=1$. Here, for interacting many-body system, the closed path is introduced through the twisted boundary and the filling $n$ can be arbitrary. Nevertheless, a remarkable feature of the Majorana phase is that, regardless of the way of introducing closed path, it can always distinguish itself from the trivial phase by producing a non-zero (topological) winding number. Such a non-zero number characterizes the topological nature of the bulk for interacting many-body systems, analogous to the role of $\pi$ Berry phase in the Kitaev chain under mean-field treatment. 

\begin{figure}[htbp]
\includegraphics[angle=0,width=65mm]{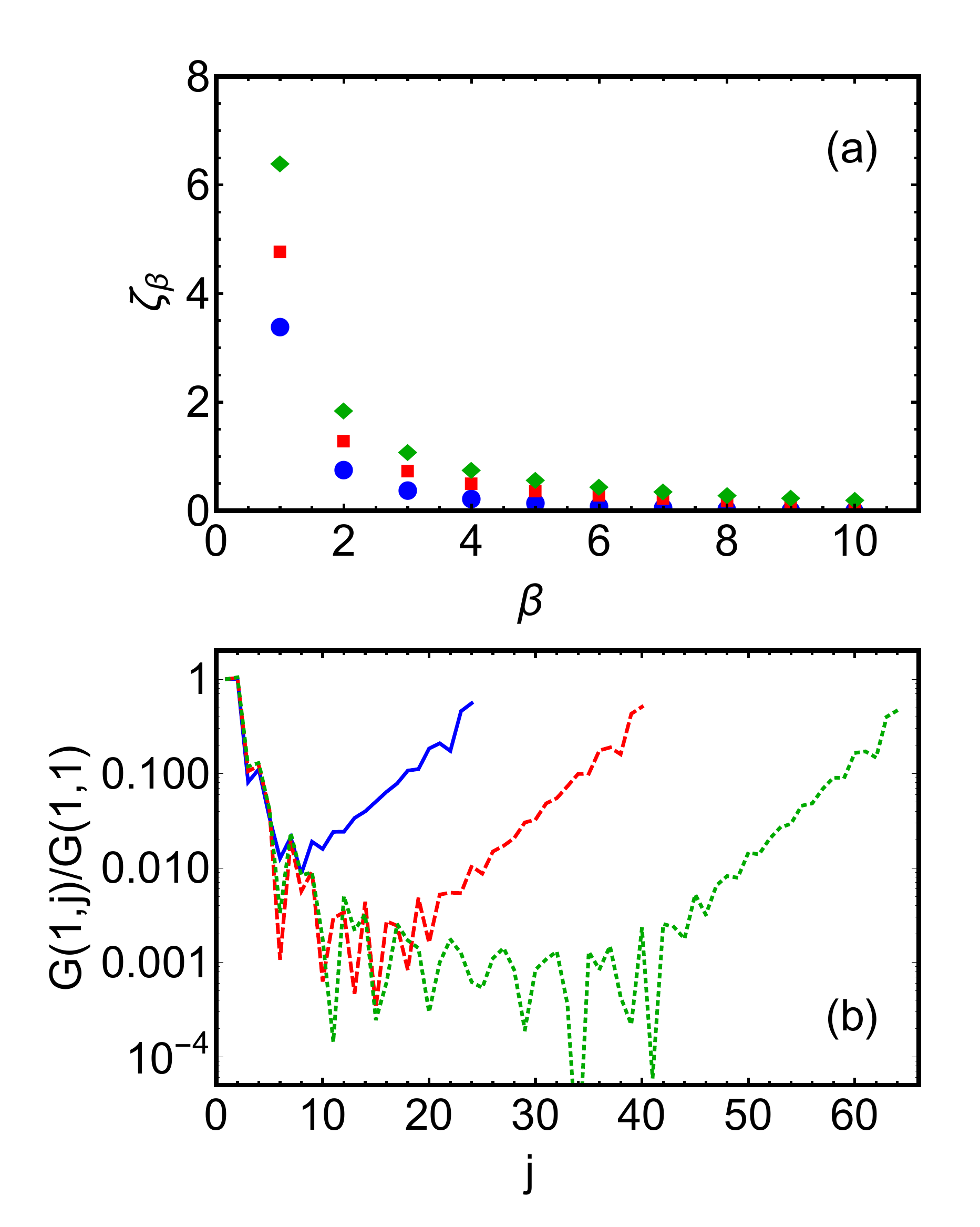}
\caption{(a) Eigenvalues $\zeta_{\beta}$ of single particle density matrix  $\rho_{ij}  \equiv \langle b^{\dagger}_{i} b_{j}\rangle$ for different system sizes $N_L=24$(Blue dots), $40$(red squares), and $64$(green diamonds), respectively. 
The eigenvalues are sorted from the largest to the smallest. 
(b) Correlation function $G(1, j)$ in log scale  for $N_L=24$(Blue dots), $40$(red squares), and $64$(green diamonds), respectively. In both panels, $n=0.75$, $g=1$, $\nu=-2$, and the system stays in strong Majorana (SM) phase.
}
\label{fig_size}
\end{figure}

Finally, we study the robustness of the Bose condensation and the edge-edge correlation against increasing the system size. We diagonalize the boson single-particle density matrix $\rho_{ij}$ for a typical SM state with different $N_{L}= 24$, $40$, and $64$ in the SM phase. Fig.~\ref{fig_size}(a) shows that in all three cases, $\rho$ has a distinct largest eigenvalue.
We also show the edge-edge correlation function $G(1, N_L)$ for different $N_L$ in Fig.~\ref{fig_size}(b), and the same strong revival at the edge is found in for all lengths studied.

\subsection{Comparison with the single-channel model}  \label{single_channel}

It is useful to contrast the Hamiltonian (\ref{eq_ham_full}) with the single-channel fermion model,
\begin{equation}
H_{sc}=\sum_{j}\left[ -t_f  (f_j^\dagger f_{j+1}+\text{h.c.} )+ UN_{f,i}N_{f,j+1} \right], \label{single_H}
\end{equation}
where $N_{f,i}= f_i^\dag f_i$ is the fermion number operator at site $i$ and $U<0$ is the attraction between neighboring-site fermions. In this model, the fermion number $N_f=\sum_i N_{f,i}$ is conserved, unlike in the resonance model ($H$ in Eq.~(\ref{eq_ham_full})). Yet this model has the same mean field theory as the resonance model, with the mean field gap defined as $\Delta/2\equiv U\langle f_i f_{i+1}\rangle $. 
 This raises the question of whether the single channel model $H_{sc}$ will also have a Majorana ground state in certain parameter regime.

To compare the ground state of single-channel model $H_{sc}$ [Eq.~(\ref{single_H})] with that of the resonance model $H$ [Eq.~(\ref{eq_ham_full}))], we shall choose the parameters ($\{U,n_{f}\}$ in $H_{sc}$ and $\{\nu,\ n\}$ in $H$) such that both models have the same fermion density $n_{f}$ and the same mean field gap $\Delta$.  With this correspondence, we have calculated the ground state of $H_{sc}$ with OBC using DMRG.

\begin{figure}[htbp]
\includegraphics[angle=0,width=65mm]{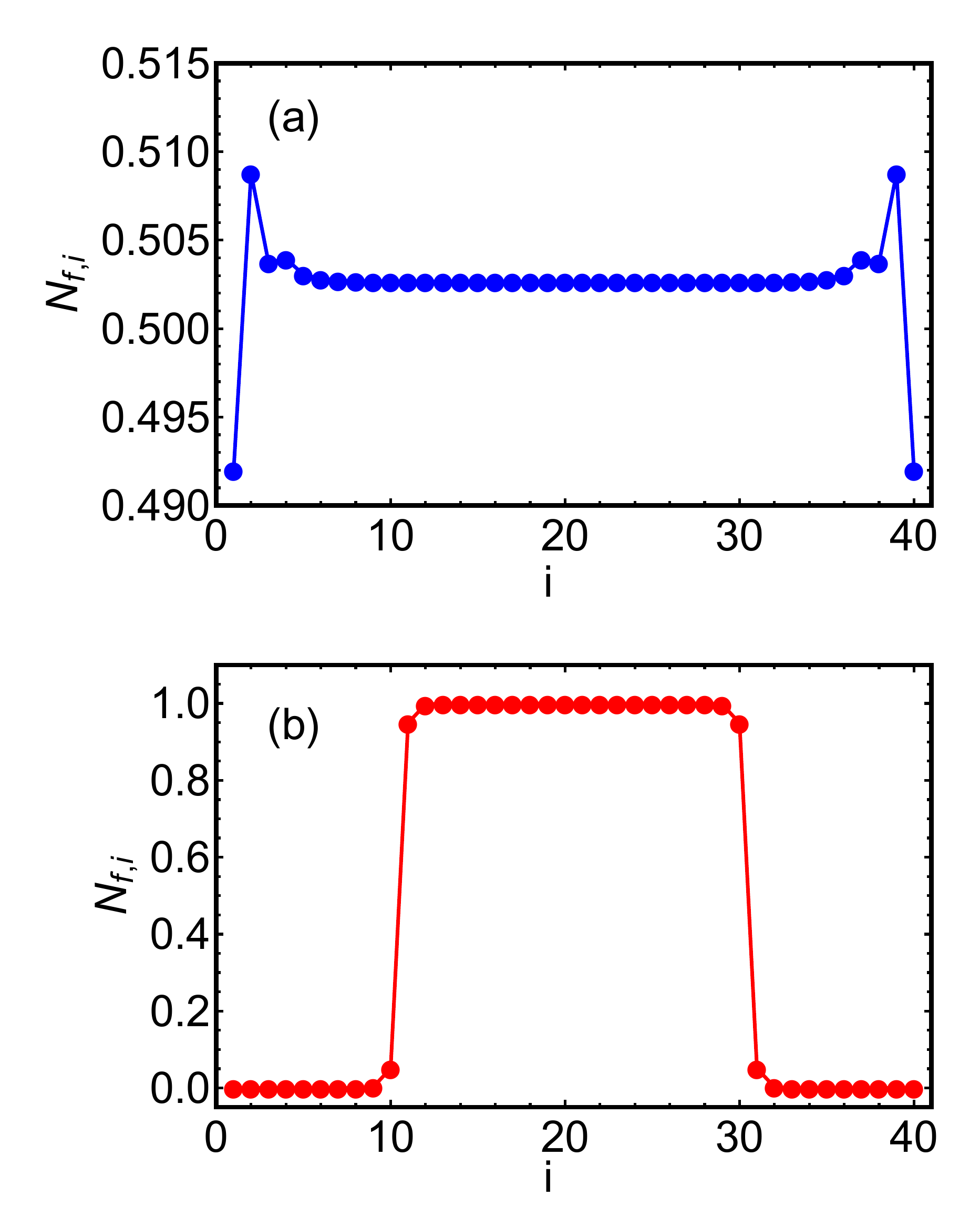}
\caption{(a) and (b) show the fermion density distributions from DMRG calculations for the  $p$-wave resonance model, Eq.~(\ref{eq_ham_full}), and  the single channel model, Eq.~(\ref{single_H}), at the same filling $n_f=0.5$ and the same mean-field gap $\Delta=2.44$. In our $p$-wave resonance model, $\nu=0.5$ and $n=1.25$ (lies inside the SM region in Fig.~\ref{fig1}). In the single-channel model, $U=-4.66$. Here $N_L=40$.}
\label{fig_comparison}
\end{figure}

We find that for all the detunings  $\nu$ in Fig.~\ref{fig1} that cover  the $SM$ and $FPP$ phases, the corresponding $U$ in $H_{sc}$ is so negative that the ground state is a droplet with all fermions packed together in a region of the size of $N_{f}$ sites, see Fig.~\ref{fig_comparison}(b). Such cluster bound state was also shown previously for few particles\cite{Berciu}. 
It can be understood by mapping $H_{sc}$ into a quantum spin chain using the Jordan-Wigner transformation,  where the occupied (empty) site is mapped to spin-up (spin-down), and the $U(<0)$ term in Eq.~(\ref{single_H}) can be mapped to the ferromagnetic Ising interaction. 
 It is then obvious that for large and negative $U$, the system forms ferromagnetic domains in the ground state, i.e., the occupied and empty sites are spatially well separated as shown in Fig.~\ref{fig_comparison}(b). Similar ferromagnetic correlation has also been shown in other 1D systems with p-wave attraction~\cite{Cui3, Gora2}.
In our calculations, the droplet may appear in different locations, as the energy difference between droplets at different locations is so small that is below our accuracy of our calculation. Clearly, the droplet phase is not the Majorana phase as found in the resonance model $H$, which exhibits the fermion density distribution as shown in Fig.~\ref{fig_comparison}(a).

For weaker attraction $U$, corresponding to $WM$ or $FHP$ regions in the large detuning limit in Fig.~\ref{fig1}, the droplet gives way to a gas phase that covers the entire chain, but still there is no strong edge-edge correlations. To conclude, the single-channel model $H_{sc}$ cannot host strong Majorana character for all couplings $U$, distinct from the SM phase in Fig.~\ref{fig1} of the resonance model.   It is also clear from 
Fig.~\ref{fig1} that in order to obtain strong Majorana correlations, the detuning $\nu$ should stay in a finite region near the two-channel resonance ($\nu\sim 0$), i.e., when bosons and fermions have comparable proportions and their conversion (or number fluctuation of fermion pairs) is the strongest. 

Here we should also remark that to describe $p$-wave Fermi gas in cold atoms experiments, the two-channel model is more realistic than the single channel model. This is because the $p$-wave resonance in these systems are generally very narrow, and the closed-channel bosons can take a sizable proportion as measured in a 3D gas near a $p$-wave resonance\cite{Toronto_K40}.

\subsection{Suppressed atom loss in the strong Majorana phase} \label{atom_loss}

Experimental realization and detection of Majorana edge state require low atom loss. For $p$-wave fermions in the lattice configuration, a previous study showed that the lattice setup will help to reduce inelastic collisional losses compared to free space~\cite{Gora}. The analysis was based on a single-channel model, and the reduced loss can be attributed to the low probability of finding three fermions close to each other outside the lattice sites~\cite{Gora}. For the present $p$-wave system described by the two-channel lowest-band model, three-fermion collision can be effectively ruled out, while the atom loss is dominantly caused by the fermion-boson or boson-boson collision at the same site. Indeed, previous studies on a continuum gas have shown that the three-body and the four-body loss rates are respectively proportional to the probabilities of finding atom-dimer and dimer-dimer at the same location~\cite{Petrov, Levinsen}, up to a background constant that is determined by the loss rate far from resonance regime.
Here, accordingly we examine the probabilities of finding a pair of boson-fermion and boson-boson at the same site, respectively denoted by $P_{bf}$ and $P_{bb}$:
\begin{eqnarray}
P_{bf}&=&\frac{1}{N_L}\sum_i \langle N_{b,i} N_{f,i} \rangle,\\
P_{bb}&=&\frac{1}{2N_L}\sum_i \langle N_{b,i}(N_{b,i}-1)\rangle,
\end{eqnarray}
with $N_{b,i}= b_i^\dag b_i$ and $N_{f,i}= f_i^\dag f_i$.

\begin{figure}[htbp]
\includegraphics[angle=0,width=85mm]{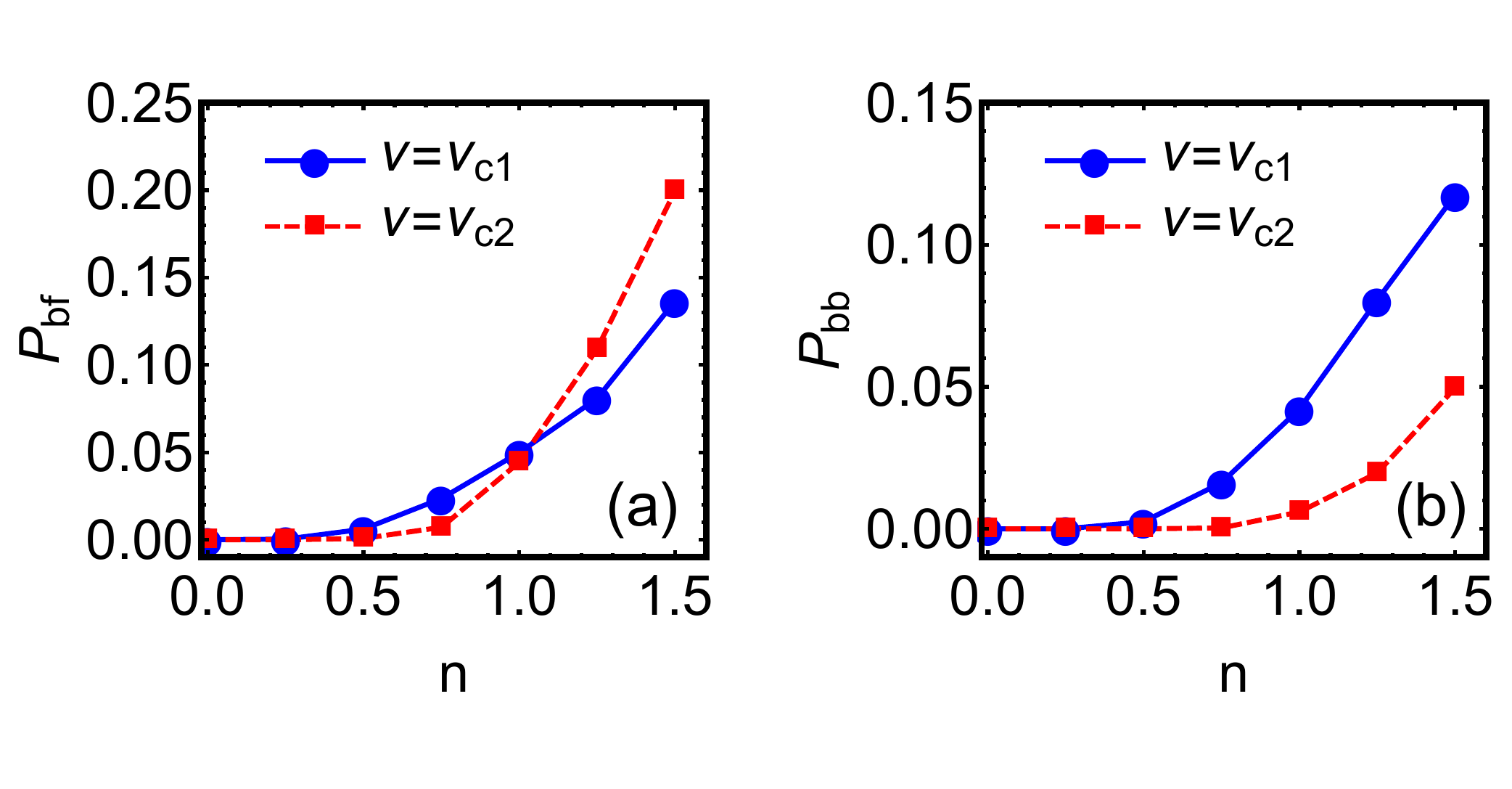}
\caption{$P_{bf}$ (a) and $P_{bb}$ (b) as functions of filling $n$ for $\nu$ staying in the lower ($\nu=\nu_{c1}$) and upper ($\nu=\nu_{c2}$) boundaries of the SM phase in Fig.~\ref{fig1}. Here, $g=1$ and $N_L=64$. }
\label{fig_loss}
\end{figure}

In Fig.~\ref{fig_loss}, we show $P_{bf}$ and $P_{bb}$ as functions of filling $n$ for $\nu$ staying in the lower ($\nu_{c1}$) and upper ($\nu_{c2}$) boundaries of the SM phase in Fig.~\ref{fig1}. We can see that for $n\lesssim 1.25$, both probabilities are less than $10\%$, suggesting the atom loss is well controlled with little atom-dimer  and dimer-dimer collisions. 
The physical reason for these low probabilities is because these configurations do not effectively take advantage of the conversion between boson and fermions ($g$-term in Eq.~\ref{eq_ham_full}) to lower the energy. 
For example, a boson and a fermion on the same site will stop the boson to convert into a fermion pair due to Pauli blocking. Similarly, if two bosons are at the same site, they cannot both convert to fermion pairs. As a result, in general the ground state does not favor the double occupations of boson-fermion or boson-boson at the same site.
However, for large fillings, such double occupations are inevitable, as shown by the increasing $P_{bf}$ and $P_{bb}$ with $n$ in Fig.~\ref{fig_loss}. These suggest that the SM phase in Fig.~\ref{fig1} should be stable enough for lower fillings ($n\lesssim 1.25$).

Now we give an estimation to the loss rate of $^{40}$K and $^{6}$Li fermions in 1D lattices. For $^{40}$K and $^{6}$Li away from Feshbach resonance, the 3D recombination rates are $\alpha_{rec}^{3D}=10^{-25} {\rm cm}^6/s$~\cite{JILA_K40} and $10^{-24} {\rm cm}^6/s$~\cite{Li6_1}, respectively. For typical transverse confinement length $\sim 50$nm and typical 1D density $\sim 10^{4}{\rm cm}^{-1}$, this leads to the decay time about 1s for $^{40}$K~\cite{Gora2} and 0.1s for $^{6}$Li when the system is out of the resonance regime (non-interacting limit). In the present case, due to the small probability $P_{bf}\lesssim 10\%$ (for filling less than unity), the actual loss rate will be further reduced by one order of magnitude, i.e, the decay time can extend to 10s and 1s, respectively, for  $^{40}$K and $^{6}$Li systems. Considering the typical hopping strength $t_f$ about tens to hundreds of Hertz, the time scale for developing the many-body correlation is a few to tens of milliseconds, much shorter than the decay time. We thus expect the Majorana phase can be observed well before severe losses occur in practical cold atoms experiment.

\subsection{Effect of inter-channel coupling} \label{effect_g}

\begin{figure}[h]
\includegraphics[angle=0,width=65mm]{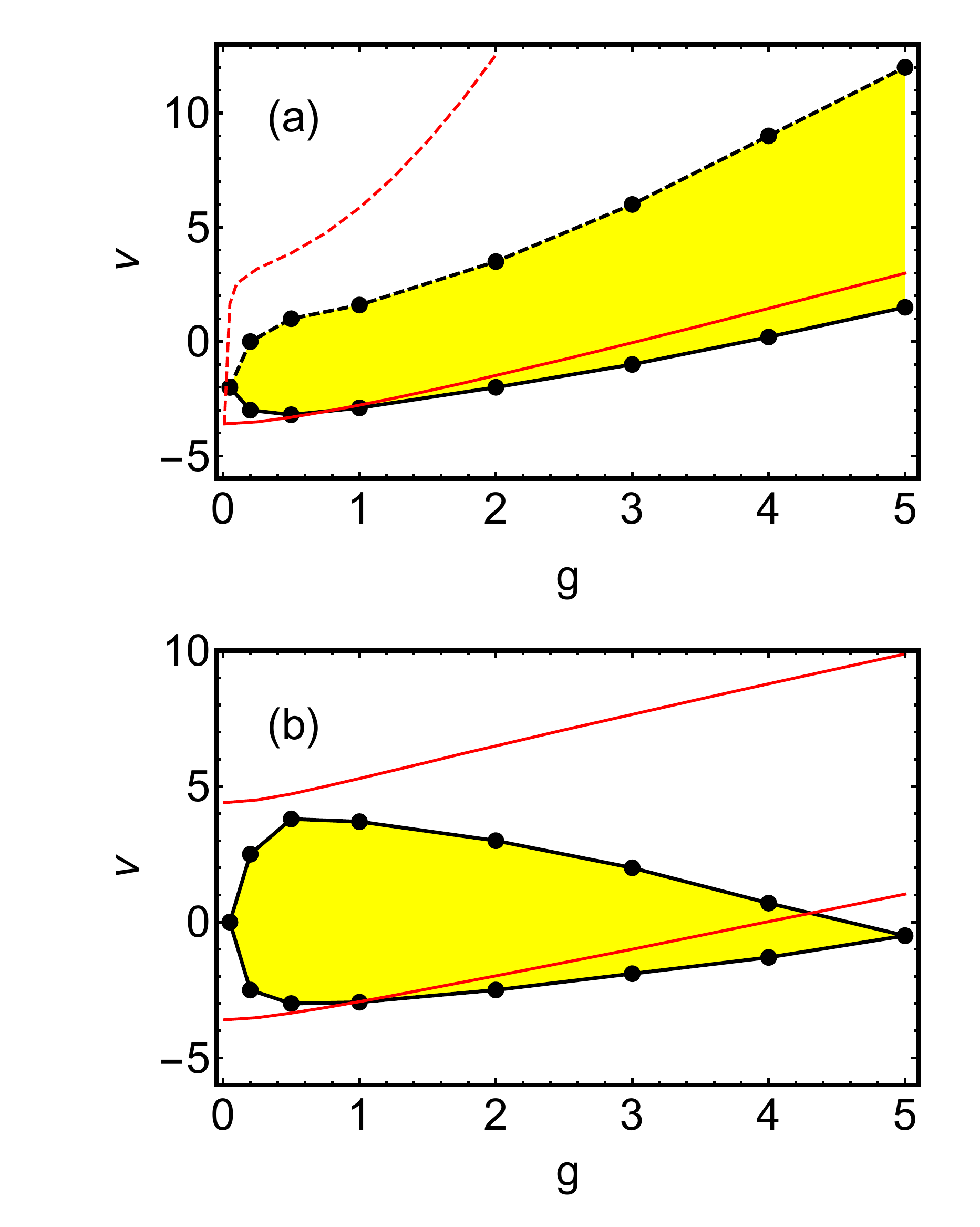}
\caption{ Phase diagram in the $g-\nu$ plane for two different fillings $n=0.75$ (a) and $1.25$ (b). Dots and black lines are from DMRG simulations after the finite-size scaling while the red lines are from mean field predictions.
The solid and dashed lines respectively mark the phase transition and the crossover.  The highlighted yellow region denotes the strong Majorana (SM) phase using the same criterion as in Fig.~\ref{fig1}.}
\label{fig_g}
\end{figure}

The phase diagram shown in Fig.~\ref{fig1} is for coupling $g=1$ (in units of hopping $t_f$). To illustrate the situation for different $g$, we have worked out the phase diagrams in  the $(g,\nu)$-plane for two different fillings $n=0.75$ and $1.25$ using DMRG.  These  results together with the mean field predictions are shown in Fig.~\ref{fig_g}.

From Fig.~\ref{fig_g}, one sees that the difference between  DMRG and mean field results grows with increasing $g$, and in large $g$ limit the mean-field theory significantly overestimate the SM region (marked by yellow color) compared to DMRG result. This can be attributed to the enhanced quantum fluctuations as increasing $g$. For filling $n<1$(Fig.~\ref{fig_g}(a)), the SM phase can always survive in a finite detuning regime, while the lower and upper boundaries both turn upward to higher detunings. In comparison, for filling $n>1$ (Fig.~\ref{fig_g}(b)), the SM phase finally disappears at a large $g_c$ ($g_c=5$ for $n=1.25)$.  For $g>g_c$ the DMRG result suggests that Majorana physics is overwhelmed by certain density waves of bosons and fermions in lattices.

Now we show that it is realistic in experiments to reach the parameter regions $(g,\nu)$ of the Majorana phase.  To give an example, we shall consider the 1D $^{40}$K Fermi gas in a lattice with depth $v\equiv V_0/E_L=6$ ($V_0$ is the lattice depth and $E_L=k_L^2/(2m)$ is the recoil energy). The hopping $t_f$ in the lowest band is  $t_f/E_L\sim 0.06$.  As shown in Ref.~\cite{Cui},  $(g,\nu)$ can be expressed by $g= g_{eff} C d^{-3/2}$, $\nu=-2g^{2}/U_{eff}$, where $g_{eff}$ is related to the effective range $r_{eff}=(mg_{eff})^{-2}$, $U_{eff}$ is the effective coupling between fermions (see Eqs.~(11) and (13) in Ref.~\cite{Cui}), and $C$ is a constant given by the overlap of Wannier functions ($C=0.06$ for $v=6$, see Fig.~4 in Ref.~\cite{Cui}). Let us consider the regime nearby the first Bloch-wave resonance, where $(l_ok_L)^{-1}\le2$ ($l_o$ is the odd-wave scattering length and $k_L=\pi/d$ is the recoil energy). The range of $U_{eff}$ and $r_{eff}$ are shown in Fig.~3 in Ref.~\cite{Cui}, from which one can estimate the range of $(g,\nu)$ in unit of $t_f$. For instance, in the interaction regime of interest, $r_{eff}$ can range from $0.75d$ to $1.5d$, so the ratio between the coupling $g$ and the hopping $t_f$ can range from $0.15$ to $0.25$. Similarly, from the information of $U_{eff}$ one can estimate the range of $\nu/t_f$ as from $-7$ to $42$. Such a broad range of $\nu/t_f$ is facilitated by the small value of $t_f/E_L$, and it well covers the SM region shown in Fig.~\ref{fig_g} for $g/t_f\in[0.15,0.25]$. Therefore the strong Majorana correlation can be achieved in a lattice with $v=6$ near a  Bloch-wave resonance.

\section{Summary and discussion}

In summary, we  have shown that the spinless Fermi gas in a 1D optical lattice near a $p$-wave resonance can have  Majorana ground state over a sizable  range of parameter space that are experimentally accessible. Our scheme makes use of the intrinsic property of cold atoms with double channels and requires neither spin-orbit coupling nor multi-chain setup. 
Our work, together with other multi-chain studies, show the number fluctuation of fermion pairs are crucial for the formation of Majorana phase. In comparison, we demonstrate that the single-channel fermions with neighboring-site attraction have no strong Majorana features. 

In identifying the phase boundaries between the Majorana phase and other trivial superfluid phases, we have examined a number of different physical quantities, including the entanglement entropy, the boson fraction and edge-edge correlation, which give rise to  consistent results as shown in Fig.\ref{fig1}. In the practical detection of Majorana phase, the low probability of  dimer-fermion and dimer-dimer pair at the same site will help to reduce atom loss.  In addition, it is proposed to identify various phases in the present system from the momentum distribution of fermions, which can be easily measured  experimentally.  Our results can be directly tested in the 1D cold atomic gases of $^{40}$K or $^{6}$Li fermions.

Finally, we further summarize our characterization of Majorana edge state in interacting many-body systems. In this work, we show that the phases that we labeled to be Majorana  (SM phase) exhibit the following  properties identical to the (number non-conserving) Kiteav chain: 
(i) The ground state in an open chain exhibits a strong edge-edge correlation that is 
robust against various kind of disorder. 
(iii) The corresponding  ground state in the bulk has a non-zero  winding number, distinguished from the nearby phases which has zero winding number. 
(iii) The phase diagram of our number-conserving Majorana phase is remarkably closed to that of the mean-field theory, which is the Kitaev model (see Fig.~\ref{fig1}). 
(iv) The properties of our Majorana state can also be interpreted from the Bosonalization method, which has been carried out in Ref.\cite{Kane} for a similar boson-fermion model.  A second-order topological transition was found, consistent with our findings as shown in Fig.~\ref{fig1}.

All above evidences (i)-(iv) show that much of the essential physics of Majorana state exhibited in the number non-conserving Kitaev chain also appear in our number conserving model.  
In other studies of Majorana physics in number-conserving models~\cite{Zoller,Diehl,Buchler,Iemini}, a ground state degeneracy between different number parity sectors has been established. Our system corresponds to 
one of the fixed number parity states (i.e. either odd or even fermion number) and we have focused on the Majorana correlation function. We shall explore that the physical effects related to the long range coherence of the Majorana correlation in future studies.

\bigskip

{\it Acknowledgment.} We would like to thank Marcello Dalmonte, Wei Yi, Wei Zhang and Miguel Cazallia for helpful discussions. This work is supported by the National Key Research and Development Program of China (2018YFA0307601, 2016YFA0300603) and the  Natural Science Foundation of China (No.11622436, No.11374177, No. 11421092, No. 11534014) awarded to X.C. and by  NSF Grant DMR-0907366, the MURI Grant FP054294-D,  the NASA Grant on Fundamental Physics 1541824, and NSFC grant (No. 11674192) awarded to T.L.H.


\begin{thebibliography}{99}

\bibitem{Majorana} E. Majorana, Nuovo Cimento {\bf 5}, 171 (1937).

\bibitem{RMP1} C. Nayak, S. H. Simon, A. Stern, M. Freedman, and S. Das Sarma, Rev. Mod. Phys. {\bf 80}, 1083 (2008).

\bibitem{Alicea} J. Alicea, Rep. Prog. Phys. {\bf 75}, 076501 (2012).

\bibitem{Kitaev} A. Y. Kitaev, Phys. Usp. {\bf 44}, 131 (2001).

\bibitem{Oreg} Y. Oreg, G. Refael, and F. von Oppen, Phys. Rev. Lett. {\bf 105}, 177002 (2010).

\bibitem{Lutchyn} R. M. Lutchyn, J. D. Sau, and S. Das Sarma, Phys. Rev. Lett. {\bf 105}, 077001 (2010).

\bibitem{Jiang} L. Jiang, T. Kitagawa, J. Alicea, A. R. Akhmerov, D. Pekker, G. Refael, J. I. Cirac, E. Demler, M. D. Lukin, and P. Zoller, Phys. Rev. Lett. {\bf 106}, 220402 (2011).

\bibitem{Liu} X.-J. Liu and H. Hu, Phys. Rev. A {\bf 85}, 033622 (2012).

\bibitem{Wei} R. Wei and E. J. Mueller, Phys. Rev. A. {\bf 86}, 063604 (2012).

\bibitem{Qu} C. Qu, M. Gong, and C. Zhang, Phys. Rev. A {\bf 89}, 053618 (2014).

\bibitem{Chen} H. Hu, C. Cheng, Y. Wang, H.-G. Luo, and S. Chen, arXiv: 1511.01762.

\bibitem{Wang}S. Wang, J.-S. Pan, X. Cui, W. Zhang, W. Yi, Phys. Rev. A {\bf 95}, 043634 (2017).

\bibitem{Moore} G. Moore and N. Read, Nucl. Phys. B {\bf 360}, 362 (1991).

\bibitem{Nayak} C. Nayak and F. Wilczek, Nucl. Phys. B {\bf 479}, 529 (1996).

\bibitem{Kitaev2}  A.Y. Kitaev,  Ann Phys. (NY)  {\bf 4321}, 2 (2006).

\bibitem{MCheng} M. Cheng and H.-H. Tu, Phys. Rev. B {\bf 84}, 094503 (2011).

\bibitem{Das_Sarma} J. D. Sau, B. I. Halperin, K. Flensberg, and S. Das Sarma, Phys. Rev. B {\bf 84}, 144509 (2011).

\bibitem{Fisher}  L. Fidkowski, R. M. Lutchyn, C. Nayak, and M. P. A. Fisher, Phys. Rev. B {\bf 84}, 195436 (2011).

\bibitem{Zoller} C. V. Kraus, M. Dalmonte, M. A. Baranov, A.M. Lauchli, and P. Zoller, Phys. Rev. Lett. {\bf 111} 173004 (2013)

\bibitem{Diehl} F. Iemini, L. Mazza, D. Rossini, R. Fazio, and S. Diehl, Phys. Rev. Lett. {\bf 115}, 156402 (2015)

\bibitem{Buchler} N. Lang and H. P. Buchler, Phys. Rev. B {\bf 92}, 041118(R) (2015)

\bibitem{Altman}J. Ruhman, E. Berg, and E. Altman, Phys. Rev. Lett. {\bf 114}, 100401 (2015).

\bibitem{Iemini} F. Iemini, L. Mazza, L. Fallani, P. Zoller, R. Fazio, and M. Dalmonte, Phys. Rev. Lett. {\bf 118}, 200404 (2017).

\bibitem{Cui} X. Cui, Phys. Rev. A {\bf 95}, 041601 (2017).

\bibitem{Sylvain} S. Nascimbene, J. Phys. B: At. Mol. Opt. Phys. {\bf 46}, 134005 (2013).

\bibitem{JILA_K40} C. A. Regal, C. Ticknor, J. L. Bohn, and D. S. Jin, Phys. Rev. Lett. {\bf 90}, 053201 (2003).

\bibitem{Li6_1}J. Zhang, E. G. M. van Kempen, T. Bourdel, L. Khaykovich, J. Cubizolles, F. Chevy, M. Teichmann, L. Tarruell, S. J. J. M. F. Kokkelmans, and C. Salomon, Phys. Rev. A {\bf 70}, 030702 (R)(2004).

\bibitem{Li6_2}
C. H. Schunck, M. W. Zwierlein, C. A. Stan, S. M. F. Raupach, W. Ketterle, A. Simoni, E. Tiesinga, C. J. Williams, and P. S. Julienne, Phys. Rev. A {\bf 71}, 045601 (2005).

\bibitem{Toronto_K40}
C. Luciuk, S. Trotzky, S. Smale, Z. Yu, S. Zhang, J. H. Thywissen, Nature Physics {\bf 12}, 599 (2016).

\bibitem{Cui1}X. Cui, Phys. Rev. A {\bf 94}, 043636 (2016).

\bibitem{Cui2}L. Zhou, X. Cui, Phys. Rev. A {\bf 96}, 030701(R) (2017).

\bibitem{footnote_model}Here we consider the physics nearby one Bloch-wave resonance, i.e., we drop the boson level index $N$ as shown in Ref.~\cite{Cui}.

\bibitem{alps1} B. Bauer et al. (ALPS Collaboration). J. Stat. Mech., 2011(05):P05001, (2011).

\bibitem{alps2} M. Dolfi et. al. Computer Physics Communications, {\bf 185}(12):3430–3440, (2014).

\bibitem{amico} Luigi Amico, Rosario Fazio, Andreas Osterloh, and Vlatko Vedral, Rev. Mod. Phys. {\bf 80}, 517 (2008).

\bibitem{PO}  O. Penrose, Philos. Mag. 42, 1373  1951;  O. Penrose and L. Onsager, Phys. Rev. {\bf 104}, 576, (1956).

\bibitem{Yang}  C. N. Yang, Rev. Mod. Phys. {\bf 34}, 694, (1962).

\bibitem{Berciu}M. Berciu, Phys. Rev. Lett. {\bf 107}, 246403 (2011).

\bibitem{Cui3}L. Yang, X.-W. Guan, X. Cui, Phys. Rev. A {\bf 93}, 051605 (R) (2016). 
\bibitem{Gora2}Y. Jiang, D. V. Kurlov, X.-W. Guan, F. Schreck, G. V. Shlyapnikov, Phys. Rev. A {\bf 94}, 011601 (R) (2016). 

\bibitem{Gora} A. K. Fedorov, V. I. Yudson, G. V. Shlyapnikov, Phys. Rev. A {\bf 95}, 043615 (2017).

\bibitem{Petrov} J. Levinsen and D. Petrov, Euro. Phys. J. D {\bf 65}, 67 (2011).

\bibitem{Levinsen} J. Levinsen, M. M. Parish, G. M. Bruun, Phys. Rev. Lett. {\bf 115}, 125302 (2015).



\bibitem{Kane}C. L. Kane, A. Stern, B. I. Halperin, Phys. Rev. X {\bf 7}, 031009 (2017).

\end{thebibliography}
\end{document}